# Vitamin-V: Expanding Open-Source RISC-V Cloud Environments

R. Canal[1*], S. Di Carlo[2], D. Gizopoulos[3], A. Scionti[4], F. Lubrano[4], J. L. Berral[1,5],
A. Call[5], D. Marron[5], K. Nikas[6], D. Pnevmatikatos[6], D. Raho[7], A. Rigo[7],
I. Papaefstathiou[8], J.M. Arnau[9], A. Arelakis[10]

[1]Universitat Politècnica de Catalunya, Spain; [2]Politecnico di Torino, Italy; [3]University of Athens, Greece; [4]LINKS Foundation, Italy; [5]Barcelona Supercomputing Center, Spain; [6]Institute of Communication & Computer Systems, Greece; [7]Virtual Open Systems, France; [8]Exapsys, Greece; [9]Semidynamics, Spain; [10]ZeroPoint Technologies AB, Sweden;

**Abstract**

*Among the key contributions of Vitamin-V (2023-2025 Horizon Europe project), we develop a complete RISC-V open-source software stack for cloud services with comparable performance to the cloud-dominant x86 counterpart. In this paper, we detail the software suites and applications ported plus the three cloud setups under evaluation.*

## Introduction

Different RISC-V processor implementations support different subsets of the ISA, virtualization, and memory hierarchies [1]. Despite recent news of RISC-V cloud deployments [2], many challenges must be addressed before RISC-V can be widely adopted for cloud applications.

**Ecosystem maturity:** the ecosystem around RISC-V processors is still developing. This includes hardware and software tools and the number of vendors and support services available.

**Performance:** while RISC-V processors can offer good performance, they may not yet be able to match the performance of more established architectures like x86 or ARM in specific applications.

**Compatibility:** many cloud applications are designed to run on x86 or ARM architectures and might not be compatible with RISC-V processors.

**Security:** as RISC-V processors become more widely adopted, they may become a target for security attacks. Ensuring the security of RISC-V-based cloud applications is an important challenge.

**Standardization:** while RISC-V is an open standard, there is still a need for further standardization in areas such as memory management and I/O interfaces.

Vitamin-V [3, 4] operates in this context and tries to address these challenges by deploying a complete RISC-V hardware-software stack for cloud services based on cutting-edge cloud open-source technologies and focusing on EPI cores[5, 6]. Next, we describe our work on well-stablished open-source cloud suites.

*Corresponding author: ramon.canal@upc.edu

Funded by the European Union. Views and opinions expressed are, however, those of the authors only and do not necessarily reflect those of the European Union or the HaDEA. Neither the European Union nor the granting authority can be held responsible for them. Project: Vitamin-V n. 101093062

## Open-Source Cloud Software

Vitamin-V aims to develop a complete RISC-V cloud open-source software (depicted in Figure 1) stack with comparable performance to the x86 counterpart.

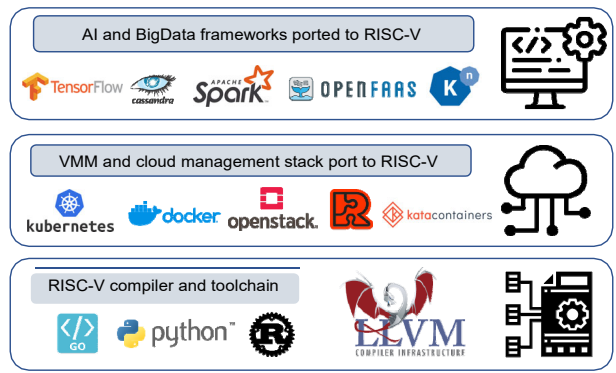

**Figure 1:** *VITAMIN-V ports of open-source suites*

Vitamin-V will deliver a complete build toolchain based on LLVM. Apart from more conventional, already supported HLLs (High-Level Languages); we will add support for GO, Python3 and Rust. Vitamin-V delivers support for validation, verification and test for cloud applications [7]. At its base, lies the hardware platform in three different flavours: QEMU, gem5 and SemiDynamic's Atrevido (deployed on AWS EC2 F1).

### Cloud setups

Vitamin-V has identified three different cloud setups (depicted in Figure 2) representing the current state-of-the-art: classic, modern and serverless.

**Classic Cloud** The Vitamin-V classic cloud software stack to be ported is based on OpenStack, one of the world's most widely deployed open-source cloud software. Porting and adaptation will consider a virtu-alization layer (the hypervisor), the host and guest OS, the runtime libraries (i.e., the Java Virtual Machine, Python3, and libc), the runtime platforms (i.e., Apache Spark for data analytics, Google TensorFlow for AI and deep learning), and benchmarking applications (i.e., TPC-DS) as an example of representative client processes running on a cloud system.

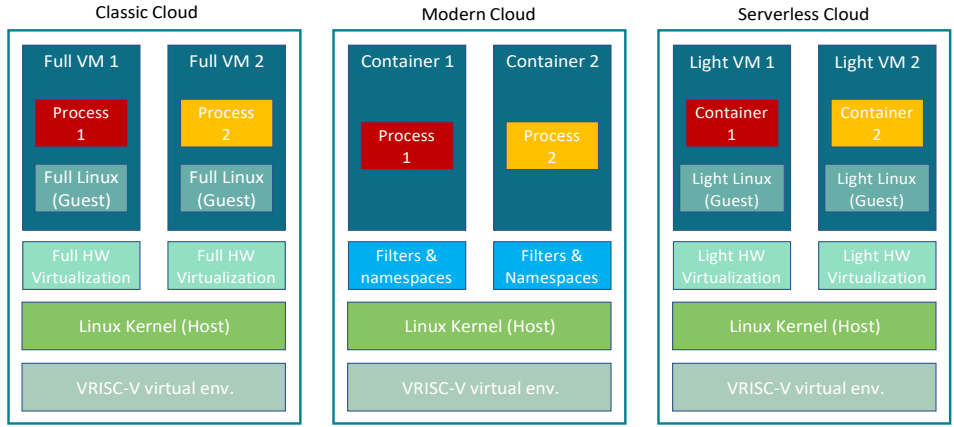

**Figure 2:** *VITAMIN-V demonstrated cloud setups*

**Modern Cloud** Moving from classic cloud architectures where full virtualization is used to modern cloud architectures requires adopting more flexible virtualization mechanisms. To this end, containerization and all related tools that automate the deployment and management of containers are of paramount importance. Kubernetes is an open-source platform for managing containerized workloads and services, developed to provide flexibility and high scalability of services. Kubernetes goes in the direction of supporting modern microservice-based cloud applications, which are composed of multiple small services interacting with each other instead of using large monolithic components.

**Serverless Cloud** Several public cloud providers move from running containers directly on the host machine to running containers inside lightweight VMs. An example of this modular architecture that runs container workloads inside VMs is Kata Containers. Kata Containers uses lightweight VMs, without relying on QEMU as a hypervisor. Alternative open-source virtual machine monitors such as Firecracker and cloud Hypervisor are written in Rust, and their building block components come from the Rust-VMM project.

## AI, Big Data, and Serverless

Application-wise, Vitamin-V will demonstrate Tensorflow (AI) and Apache Spark (Data Analytics) workloads on the three cloud setups. The three setups will be benchmarked against relevant AI applications (i.e., Google Net, ResNet, VGG19), Big-Data applications (TPC-DS on top of Apache Spark), and Serverless applications (FunctionBench, ServerlessBench). Vitamin-V aims to match the software performance of its x86 equivalent (x86 is the dominant ISA in cloud servers). Given the limited availability of RISC-V hardware, measures will be averaged by CPU core mark scores to achieve a fair assessment.

## Conclusions

In a new era of open and collaborative hardware designs, integrating RISC-V in cloud computing represents a crucial opportunity to revolutionize the industry with its unparalleled flexibility, efficiency, and scalability. Vitamin-V contributes all across the stack from the CPU, compiler, toolchain, cloud management up to applications to this challenging revolution.